# New phase transitions by changing the even number of components of ordering field near Lifshitz's point in melt-crystallized polymers

A. N. Yakunin[*]

*Karpov Institute of Physical Chemistry, 105064 Moscow, Russia*

**Abstract:** The transition from n = 0 to n = 2 is revealed where n is the number of components of ordering field. The critical exponents are estimated. In frameworks of scaling theory of phase transitions and critical phenomena the results obtained are in a good agreement with experimental and theoretical data.

About the $\alpha$ – relaxation transition the draw ratio at break, $\lambda_{br}$, has a maximum [1, 2] in copolymers of low pressure ethylene – acrylic acids with the small number of the latter comonomer (~0.1-0.3mol% [1, 3]) as well as in linear PE (see [4] and refs. in it). At this point called by the second Lifshitz point [2] the expression $\lambda_n^2/\lambda_{br}^2 = P$ (obtained for room temperature) should be changed as formulated in [2]. Here, $\lambda_n$ is the neck draw ratio, $P = l_a/l_a^{cr}$ is the probability of collision of chain ends, as a function of N $l_a$ is the mean thickness of amorphous layers in isotropic material, $l_a^{cr}$ is the same value at $N = N_{cr}$, $N = M_w/14$ is the polymerization degree, 14kg/kmol is the molecular mass of repeated $CH_2$ group, $\ln N_{cr} \approx 15.89$. Figure 1 shows the dependence of $\lambda_{br}$ on N at room [5] and elevated [4] temperatures. The theoretical curves $\lambda_{br} = (N/N_{cr})^{(1-\nu d)/2}$ are also presented in Fig. 1, $\nu$ is the critical exponent of correlation radius, d is the space dimension. It has been recently shown [6] that $\nu = 0.5(1 + t - 1/t)$ where $t = 1 + 2\pi(n+2)/p$, p = 137, n is the number of components of ordering field; it should be underlined that n may be only by the even number, otherwise, the gauge symmetry breaks. The theoretical curves in Fig. 1 seem to confirm our assumption that near the second Lifshitz point the phase transition from n = 0 to n = 2 occurs. Here, we have used precise value for $\ln N_{cr} \approx 15.69$ in comparison with [2].

In Table 1 the theoretical values of critical exponents $\nu$, $\gamma$, $\beta$ are presented. If n = 1 or n = 3 then such systems can be regarded as examples of materials when states with n = 0 and n = 2

---

[*] E-mail: yakunin@cc.nifhi.ac.ru

(or n = 2 and n = 4) must have the same probability, while n = 6 seems to be realized in some percolation systems such as gels [7]. The experimental value of the exponent η is not determined exactly. The theoretical estimations [8, 9] show its weak dependence on n. In this connection, we may attribute any reasonable meaning to η. Then for β and γ can be obtained the following evaluations by standard ways (γ + β = δβ) [8] if to suppose η = δ/p and to use the identity δ = -1 + 2d/(d − 2 + η). Thus, η ≈ 0.0350, γ = νd(δ − 1)/(δ + 1), β = νd/(δ + 1). These values are presented in Table 1. They are in accordance with recent theoretical data [9]. However, more accurate experimental checking could be made in future.

Table 1. Critical exponents to the 4th decimal place.

| n | ν | γ | β | (νd-1)/2 |
|---|---|---|---|---|
| 0 | 0.58787 | 1.15516 | 0.30423 | 0.38181 |
| 2 | 0.66923 | 1.31503 | 0.34633 | 0.50385 |
| 4 | 0.74549 | 1.46487 | 0.38579 | 0.61823 |
| 6 | 0.81766 | 1.60669 | 0.42314 | 0.72649 |
| 1 | 0.62927 | 1.23650 | 0.32565 | 0.44390 |
| 3 | 0.70793 | 1.39106 | 0.36636 | 0.56189 |

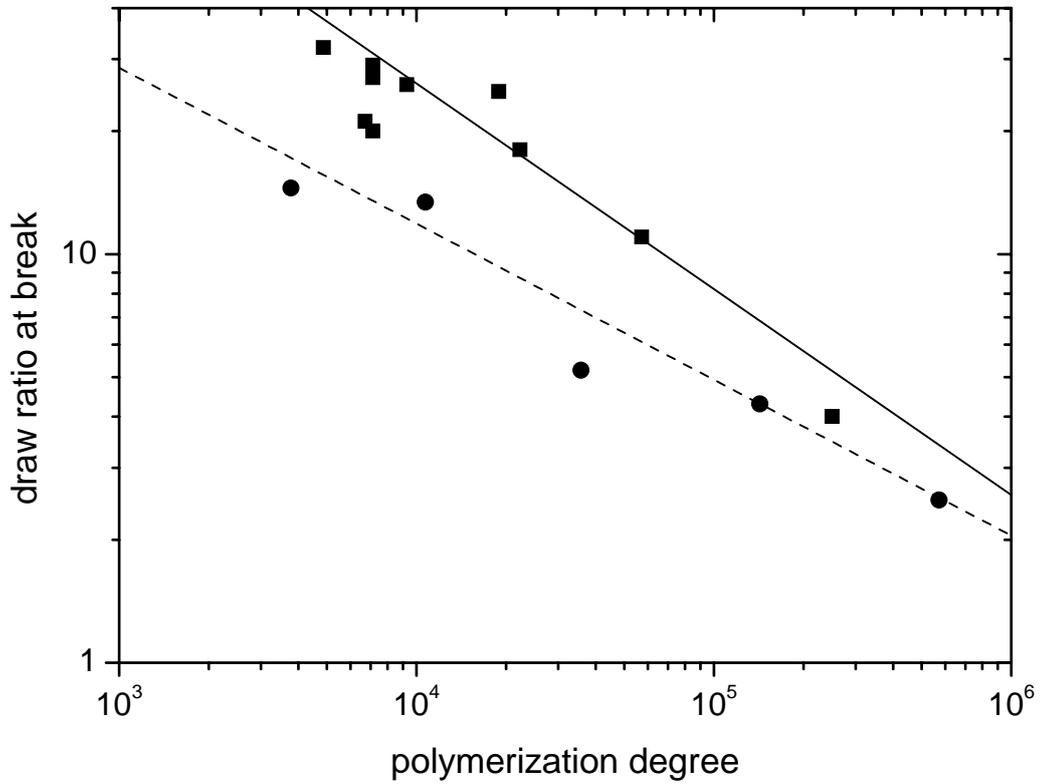

Fig. 1. The draw ratio at break, $\lambda_{br}$, *vs* the polymerization degree, N. Experimental data are obtained at room temperature [5] and at 75C [4], they are marked by circles and by squares, respectively; the theoretical curves $\lambda_{br} = \exp(-0.5(\nu d - 1)\ln(N/N_{cr}))$ correspond to $N_{cr} \approx \exp 15.69$ and $\nu$ from Table 1 for n = 0 (dash line) and n = 2 (solid line).